\documentclass[conference]{IEEEtran}
\IEEEoverridecommandlockouts
\usepackage{cite}
\usepackage{amsmath,amssymb,amsfonts}
\usepackage{algorithmic}
\usepackage{graphicx}
\usepackage{textcomp}
\usepackage{xcolor}
\def\BibTeX{{\rm B\kern-.05em{\sc i\kern-.025em b}\kern-.08em
    T\kern-.1667em\lower.7ex\hbox{E}\kern-.125emX}}

\usepackage{subfigure}
\usepackage{multirow}
\usepackage{makecell}
\usepackage{enumitem}
\setenumerate[1]{itemsep=0pt,partopsep=0pt,parsep=\parskip,topsep=0pt}
\setitemize[1]{itemsep=0pt,partopsep=0pt,parsep=\parskip,topsep=0pt}
\setdescription{itemsep=0pt,partopsep=0pt,parsep=\parskip,topsep=0pt}
\usepackage{fontawesome5}

\usepackage{hyperref}

\begin{document}

\title{The Unwritten Contract of Cloud-based\\ Elastic Solid-State Drives}

% \author{
% \IEEEauthorblockN{}
% \IEEEauthorblockA{
% \textit{}}
% \and
% \IEEEauthorblockN{}
% \IEEEauthorblockA{
% \textit{}}
% }

\author{
\IEEEauthorblockN{Yingjia Wang and Ming-Chang Yang\thanks{This work is supported in part by The Research Grants Council of Hong Kong SAR (Project No. CUHK14218522).}}
\IEEEauthorblockA{
\textit{The Chinese University of Hong Kong}}
}

% \and
% \and
% \and
% \and
% \and

\maketitle

\begin{abstract}

Elastic block storage (EBS) with the storage-compute disaggregated architecture stands as a pivotal piece in today's cloud. 
EBS furnishes users with storage capabilities through the elastic solid-state drive (ESSD).
Nevertheless, despite the widespread integration into cloud services, the absence of a thorough ESSD performance characterization raises critical doubt: when more and more services are shifted onto the cloud, can ESSD satisfactorily substitute the storage responsibilities of the local SSD and offer comparable performance?

In this paper, we for the first time target this question by characterizing two ESSDs from Amazon AWS and Alibaba Cloud.
We present an unwritten contract of cloud-based ESSDs, encapsulating four observations and five implications for cloud storage users.
Specifically, the observations are counter-intuitive and contrary to the conventional perceptions of what one would expect from the local SSD.
The implications we hope could guide users in revisiting the designs of their deployed cloud software, i.e., harnessing the distinct characteristics of ESSDs for better system performance.

\end{abstract}

% \begin{IEEEkeywords}
% solid-state drive, cloud storage, unwritten contract
% \end{IEEEkeywords}

% ------------------------------------------
\section{Introduction} \label{sec:intro}
% ------------------------------------------

% Cloud computing provides a new service paradigm with diverse hardware configurations, scalable resource management, and ease of deployment.
% Compute power/resources and services are delivered in the cloud with pay-as-you-go-procing.
% This has attracted many customers to shift their services (high-performance computing, deep learning, data analysis, and online websites) and data to the cloud for elastic, stable, reliable, and pay-as-you-go pricing.
% Cloud computing not only benefits cloud-native services but also empowers edge services that lack computing power and storage capacity~\cite{rellermeyer2013cloud}.

\textit{Elastic block storage (EBS)} has become the cornerstone of today's cloud, underpinning the high-performance storage demand from diverse cloud services~\cite{miao2022luna,li2023depth,zhang2024s,wang2024ransom,shu2024burstable}.
As shown in Figure~\ref{ebs}, EBS typically adopts a storage-compute disaggregated architecture, with compute and storage clusters located separately and interconnected by the high-speed datacenter network.
In EBS, the virtual machines (VM) for users are hosted in the compute cluster while the user data is persistently stored in the storage cluster.
This architectural innovation not only enhances the efficiency and elasticity of the cloud infrastructure, but also improves the cost-effectiveness for users on a flexible and pay-as-you-go basis.

EBS provides storage resources for users in the form of \textit{elastic solid-state drive}, which, for simplicity, is called \textit{ESSD} in the following.
ESSD is a virtualized storage device attached to the user VM, employing the block interface and supporting random access (e.g., read and write) in the storage space.
From the perspective of the user, ESSD is similar to the well-known local SSD, on which the existing filesystems and applications can be directly deployed.
But differently, the most noteworthy feature of ESSD is that it breaks the physical performance and capacity boundaries of the local SSD.
This makes it possible to elastically guarantee performance level (e.g., in throughput and IOPS) and allocate storage capacity, to accommodate the ever-changing demands of users.

The rapid development of cloud computing has fostered the widespread adoption of ESSDs in numerous cloud services.
\textit{However, to the best of our knowledge, there has been no comprehensive performance characterization of cloud-based ESSDs.}
As the majority of host software now is designed and optimized for the local SSD, this research gap raises concerns in the following question: \textit{when more and more services are shifted onto the cloud, can ESSD satisfactorily substitute the storage responsibilities of the local SSD and offer comparable performance?}

To this end, we make the first endeavor to answer this question by characterizing two ESSDs from two leading cloud storage providers in the world: Amazon AWS~\cite{website:awsssd} and Alibaba Cloud~\cite{website:alissd}.
Based on extensive experiments, we surprisingly find that ESSD exhibits several counter-intuitive performance features, contrary to the common beliefs of the local SSD.

% ------------------------------------------
\vspace{.5em}
\noindent\fbox{\parbox{0.475\textwidth}{

\centering
\vspace{.5em}
\textbf{The Unwritten Contract of Cloud-based ESSDs}
\vspace{.5em}

\raggedright
\noindent\textit{\textbf{Observations}}
\begin{enumerate}[leftmargin=*]
    \item \textit{The latency of ESSDs is tens to a hundred times higher than that of SSD when I/Os are not well scaled up (i.e., I/O size is small and/or I/O queue depth is low).}
    \item \textit{The performance impact of GC appears much later or even disappears.}
    \item \textit{The throughput of random writes outperforms that of sequential writes, reaching a maximum of 1.52$\times$ and 2.79$\times$ in two ESSDs, respectively.}
    \item \textit{The maximum bandwidth is deterministic and no longer sensitive to the access pattern.}
\end{enumerate}

\vspace{.5em}

\noindent\textit{\textbf{Implications}}
\begin{enumerate}[leftmargin=*]
    \item \textit{Scale the I/O sizes and I/O queue depths up as much as possible.}  
    \item \textit{Reconsider if and how existing GC-mitigated techniques based on local SSDs should be correspondingly adapted for ESSDs.}
    \item \textit{Rethink if it is still worthwhile to convert random writes in random-write-based software into sequential writes and if it is beneficial to proactively trigger random writes in sequential-write-based software.}
    \item \textit{Smooth the read/write I/Os to be evenly distributed across the timeline and below the guaranteed throughput budget.}
    \item \textit{Re-evaluate I/O reduction techniques (e.g., compression, deduplication) that were previously considered to impair performance.}
\end{enumerate}
}

}
\vspace{.5em}
% ------------------------------------------

We present an unwritten contract of cloud-based ESSDs, organized into four observations and five implications, as shown above.
Over the past years, cloud storage users have been requesting ever-higher service-level agreements (SLA).
This places higher expectations not only on cloud storage providers but also on users themselves in terms of their deployed software.
The unwritten contract we offer in this paper could benefit cloud storage users to revisit their cloud software designs, specifically, to unleash the better potential of ESSDs by leveraging unique performance characteristics.
We open-source the evaluation framework to stimulate the following studies\footnote{\url{https://github.com/yingjia-wang/UC}}.

In the following of this paper, we first provide the background of (local) SSD, EBS, and ESSD in~\S\ref{sec:bg}.
We present the unwritten contract in detail in~\S\ref{sec:uc}.
The related work and conclusion are introduced in~\S\ref{sec:related} and~\S\ref{sec:conclusion}, respectively.

% ------------------------------------------
\section{Background} \label{sec:bg}
% ------------------------------------------

% ------------------------------------------
\subsection{Flash-based Solid-State Drive (SSD)} \label{sec:bg_ssd}
% ------------------------------------------

\begin{figure}[t]
    \centering
    \includegraphics[width=0.38\textwidth]{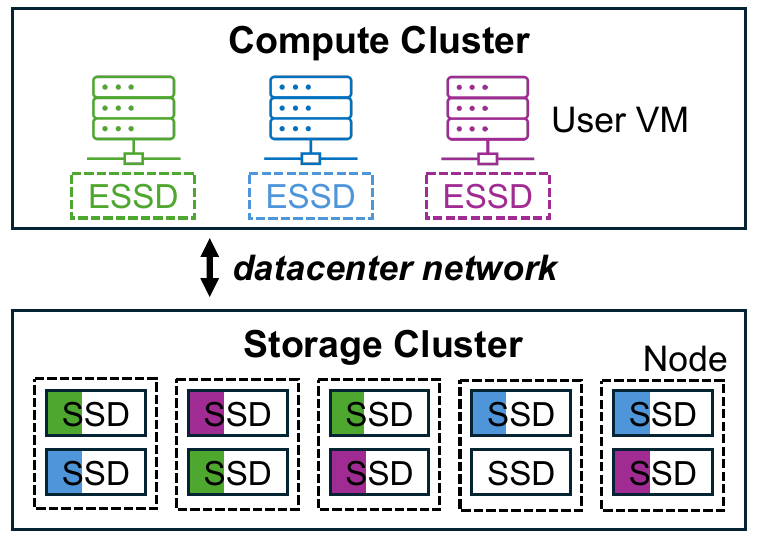}
    % \vspace{-.3em}
    \caption{
    \textbf{The storage-compute disaggregated architecture of elastic block storage (EBS).}
    ESSD is attached to the user virtual machine (VM) in the compute cluster.
    The physical storage space of an ESSD is distributed and replicated across different nodes and SSDs in the storage cluster.
    }
    % \vspace{-.5em}
    \label{ebs}
\end{figure}

\textit{Flash-based SSD} has achieved a high and continuously increasing share in the storage market due to the benefits in performance, energy efficiency, and impact resistance.
According to IDC, the byte shipment share of SSDs is expected to reach about 40\% of the total in 2025~\cite{website:idc}.
Flash-based SSD typically adopts a multi-level architecture, including channel, die, plane, block, and page.
The flash die is the minimum unit of parallel operations while the flash page is the minimum unit of data storage.
To improve SSD firmware management efficiency, flash blocks are typically grouped into superblocks, each of which comprises flash blocks across many flash dies to fully leverage flash parallelism.

To communicate with the host, the SSD longstandingly employs the \textit{block interface}, which abstracts the storage space into an array of logical blocks (e.g., 4KB) that can be randomly read or written.
While this simple abstraction has successfully contributed to the broad adoption of SSDs, a complex \textit{flash translation layer (FTL)} is required in the SSD to bridge the gap with the underlying flash characteristics (e.g., erase-before-write).
The two main duties of FTL are address mapping and garbage collection (GC).
The address mapping converts logical addresses from the host to the physical addresses in the SSD by keeping track of a fine-grained (e.g., page-level) mapping table.
The GC is carried out periodically to reclaim invalid space in the granularity of flash blocks, when the valid pages in some blocks are relocated and these blocks can be erased.
FTL is also responsible for other tasks such as wear-leveling~\cite{jiao2022wear}, error correction~\cite{zhao2013ldpc}, and bad block management~\cite{yen2022efficient}.

% To bridge the gap between the block interface and underlying flash characteristics (e.g., out-of-place update, erase-before-write), the SSD necessitates a complex flash translation layer (FTL) inside.
% Despite facilitating the widespread adoption of SSDs, the block interface has led to an ever-rising tax in performance and cost~\cite{bjorling2021zns,kang2014multi,bjorling2017lightnvm,sabol2023fdp}.
% On the one hand, FTL conducts fine-grained logical-to-physical address mapping, which requires large on-board DRAM (e.g., 4GB for a 4TB SSD) and thus considerably raises the hardware cost.
% On the other hand, FTL performs  to reclaim invalid space, which degrades SSD performance and necessities large flash over-provisioned (OP) space (e.g., up to 50\% in CacheLib as reported in~\cite{berg2020cachelib}) to counteract this effect.

% ------------------------------------------
\subsection{Elastic Block Storage (EBS)} \label{sec:bg_ebs}
% ------------------------------------------

As shown in Figure~\ref{ebs}, \textit{elastic block storage (EBS)} typically employs a storage-compute disaggregated architecture with independent compute and storage clusters, and the two clusters are not physically co-located but interconnected with the high-speed datacenter network.
% (e.g., compute cluster, partitioning cluster, and persistence cluster~\cite{wang2024ransom,shu2024burstable}).
The \textit{compute cluster} hosts virtual machines (VM) for users and forwards user requests to the backend storage cluster, while the \textit{storage cluster} persists or fetches requested data in a distributed filesystem.

The disaggregation architecture of storage and compute brings out several benefits~\cite{miao2022luna,shu2024burstable,zhang2024s}.
First, the storage and compute resources can be elastically paid and allocated according to the user target workloads.
This allows for higher resource utilization while enhancing the cost-effectiveness for users, compared to offering general-purpose servers.
Second, the resources in storage and compute clusters can be maintained and scaled independently, thereby realizing higher maintainability and scalability.
Third, it is simpler and faster to migrate application services across compute servers, given that their states can be persistently stored in storage servers.

% First, compute and storage servers can be designed independently
% such that either type of server is optimized for its target workloads, which is more cost-efficient than general-purpose servers; Second, the storage cluster is dedicated to ensuring the safety of data and manages the massive storage media with high utilization to lower the cost. 
% It becomes a standalone and common component in the cloud infrastructure that is easy to be integrated by different compute applications.

% ------------------------------------------
\subsection{Elastic Solid-State Drive (ESSD)} \label{sec:bg_essd}
% ------------------------------------------

\begin{table*}[t]
    \caption{\textbf{The configurations of two ESSDs and SSD.}
    % BW: bandwidth. QD: queue depth.
    }
    \centering
    \vspace{-.3em}
    \small
    \begin{tabular}{c|c|c|c|c|c|c}
    \hline
    & \textbf{Provider and Type} & \textbf{Max. BW (GB/s)} & \textbf{Max. IOPS} & \textbf{Cap. (TB)} & \textbf{VM Type} & \textbf{Region} \\ \hline \hline
\textit{ESSD-1} & Amazon AWS io2~\cite{website:io2} & $\sim$3.0 & 25.6K & 2 & m6in.xlarge & Tokyo \\ \hline
\textit{ESSD-2} & Alibaba Cloud PL3~\cite{website:alissd} & $\sim$1.1 & 100K & 2 & ecs.g5.4xlarge & Hangzhou \\  \hline
\textit{SSD} & Samsung 970 Pro~\cite{website:samsung970pro} & Seq. R/W 3.5/2.7 & \makecell[c]{Rand. R/W 500K/500K \\(4KB, QD32)} & 1 & / & / \\
    \hline
    \end{tabular}
    % \vspace{.5em}
    \label{essd_config}
\end{table*}

EBS provides storage resources for users in the form of \textit{elastic solid-state drive (ESSD)}, which is a virtualized storage device attached to user VM.
ESSD can be created, attached, unattached, and destroyed without disrupting the user VM.
Figure~\ref{ebs} illustrates the mapping relationship between the virtualized ESSD and the (multiple) physical SSDs.
That is, the physical storage space of an ESSD is typically distributed and replicated (e.g., three-way~\cite{zhang2024s}) across different nodes and SSDs in the storage cluster, for the purposes of load balancing and data availability.

From the user's point of view, ESSD is similar to the well-known local SSD in that ESSD also employs the block interface and supports random access in the storage space.
This ensures seamless compatibility with the existing software ecosystem, and facilitates the existing file systems and applications can be directly deployed on ESSDs.

% The ability to leverage these familiar functionalities simplifies the adoption of ESSDs in various computing environments while preserving the flexibility and performance benefits associated with this advanced storage technology.

Nonetheless, beyond the local SSD, ESSD delivers several additional benefits~\cite{zhang2024s,website:awsssd,website:alissd}.
First, ESSD is virtualized, which breaks the physical performance and capacity boundaries in the local SSD.
As a result, both the performance level (e.g., in throughput and IOPS) and the storage capacity can be flexibly requested in response to the ever-changing user requirements.
Second, ESSD can ensure high data availability and reliability, thanks to the data replication across multiple nodes to prevent single-point failure.
Third, ESSD also enables some advanced features such as encryption and snapshotting.

% (e.g., Amazon AWS~\cite{website:awsssd}, Alibaba Cloud~\cite{website:alissd}, Microsoft Azure~\cite{website:azureessd}, Google Cloud~\cite{website:googleessd})

Cloud storage providers typically offer multiple ESSD types with varying performance levels.
Different performance levels guarantee different budgets in maximum throughput and maximum IOPS (higher is usually more expensive).
For reference, Amazon AWS~\cite{website:awsssd} supports general-purposes types (i.e., gp2, gp3) and provisioned-IOPS types (i.e., io1, io2), where io2 is the most advanced one.
Alibaba Cloud~\cite{website:alissd} supports four performance levels: PL0, PL1, PL2, and PL3 (the higher the number the better the performance).

% ESSD is generally priced based on the performance level, storage capacity, and usage time.
% ESSD, similar to the user VM, is paid flexibly for its performance level, storage capacity, and others such as extra provisioned resources and network flow.
% The cloud storage providers, such as AWS, offer a variety of cost-efficient storage volumes for users to meet their distinct needs and adapt to the changing market.
% Such agility and flexibility allow enterprises to right-size storage for different workloads in the cloud.

% ------------------------------------------
\section{Unwritten Contract of Cloud-based ESSDs} \label{sec:uc}
% ------------------------------------------

% Specifically, the ESSD from AWS (i.e., ESSD-1) supports at most 1300GB/s bandwidth and up to 100K IOPS, has 2TB capacity, and runs on a VM with m6in.xlarge.
% The ESSD from Alibaba Cloud (i.e., ESSD-2) supports 1GB/s bandwidth and up to 100K IOPS, has 2TB capacity, and runs on a VM with ecs.g5.4xlarge.

Despite the prevalent integration in diverse cloud services, to the best of our knowledge, the performance features of the ESSD have not been adequately investigated and revealed.
In this section, we present an unwritten contract of cloud-based ESSDs.
In the following, we begin with the experimental setups, and then introduce our unwritten contract in detail.
% ,  incorporating four observations and five implications.

Note that this paper focuses on the device-level performance characterization as in prior work~\cite{wu2019towards,saha2021kv,doekemeijer2023performance} without exploring specific pieces of software, in the hope of benefiting a wider range of cloud storage users to revisit their software designs.

% ------------------------------------------
\subsection{Experimental Setups} \label{sec:uc_expsetup}
% ------------------------------------------

To strengthen the generalizability of findings, we characterize two ESSDs from two leading cloud providers in the world (i.e., Amazon AWS~\cite{website:awsssd} and Alibaba Cloud~\cite{website:alissd}), and the configurations of ESSDs are summarized in Table~\ref{essd_config}.
Specifically, both ESSDs have the highest-class types available from their providers, making their maximum bandwidth comparable to that of the local SSD.

To demonstrate the features of the local SSD for comparison, we also evaluate Samsung 970 Pro~\cite{website:samsung970pro}, a popular SSD in research, and its configurations are also listed in Table~\ref{essd_config}.
Experiments on Samsung 970 Pro are running on a workstation PC, equipped with Intel® Xeon® W-2245 Processor, where the operating system is 64-bit Ubuntu 20.04 LTS with Linux kernel version 5.15.0.

% research~\cite{wu2019towards,wu2021storage,oh2023cjfs}.
% In addition, the VMs where ESSDs are attached can satisfy the minimum performance requirements of the corresponding ESSDs~\cite{TODO}.

We employ the widely-used FIO benchmark tool~\cite{website:fio} to generate workloads with diverse loads (i.e., I/O size and I/O queue depth) and access patterns (i.e., random write, sequential write, random read, and sequential read).
The detailed workload settings are introduced in each subsection below.

% ------------------------------------------
\subsection{Latency Performance} \label{sec:uc_lat}
% ------------------------------------------

\begin{figure*}[t]
    \centering
    \subfigure[Average Latency of ESSD-1 from Amazon AWS]{
    \includegraphics[width=0.5\textwidth]{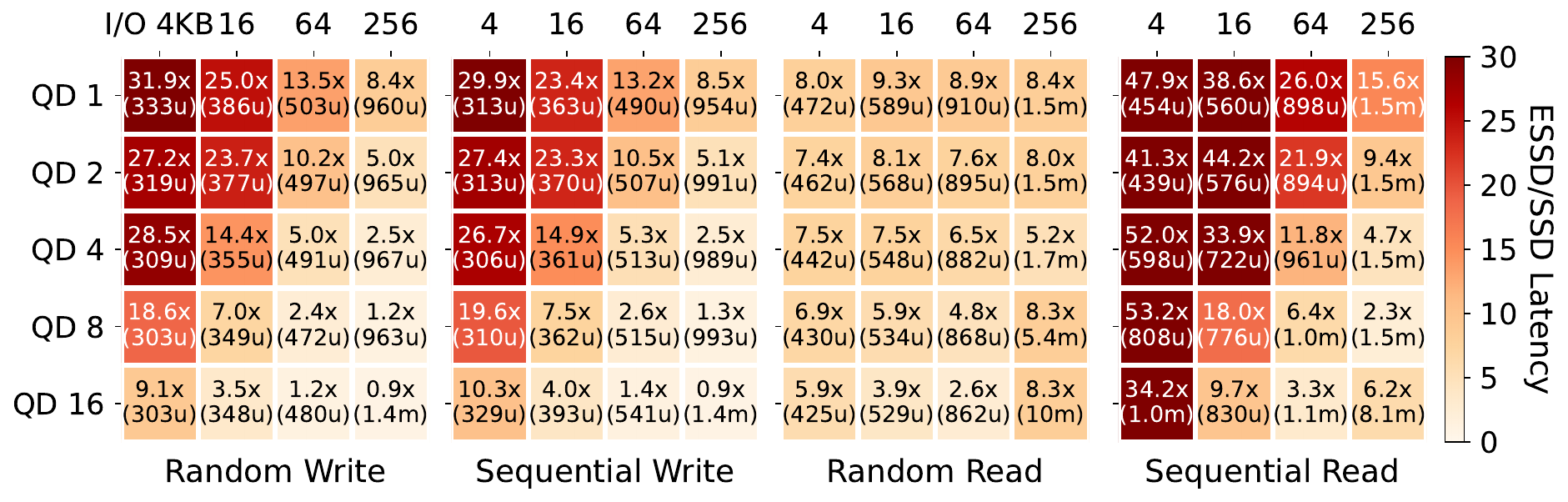}
    }
    \hspace{-2.1em}
    \subfigure[P99.9 Latency of ESSD-1 from Amazon AWS]{
    \includegraphics[width=0.5\textwidth]{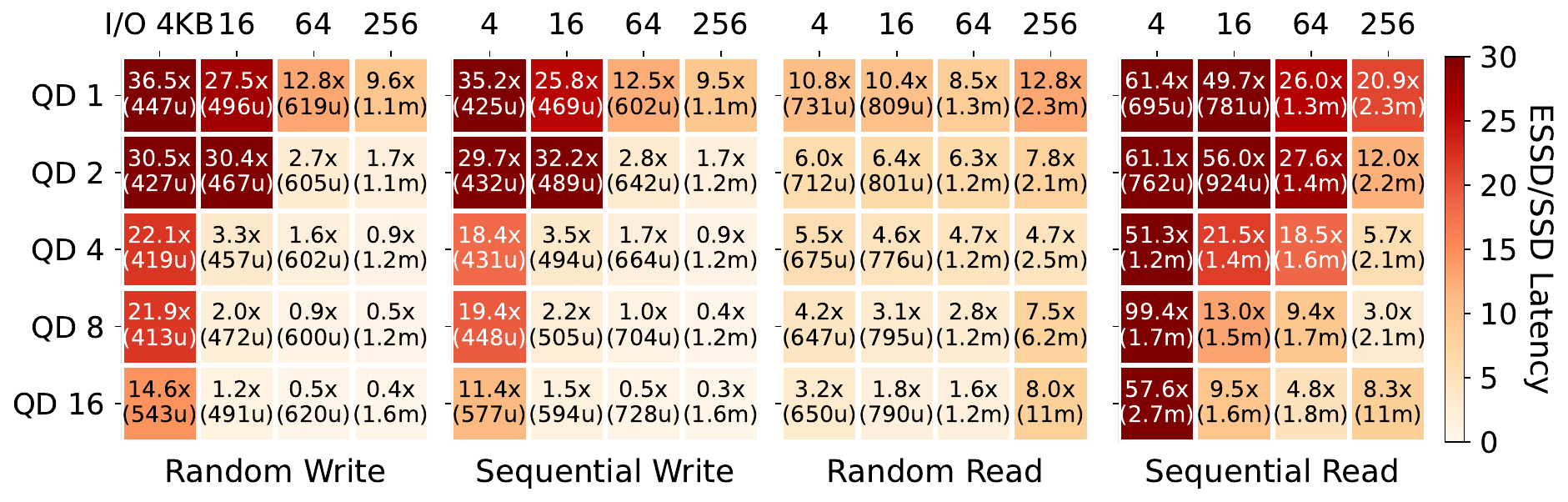}
    }
    \subfigure[Average Latency of ESSD-2 from Alibaba Cloud]{
    \includegraphics[width=0.5\textwidth]{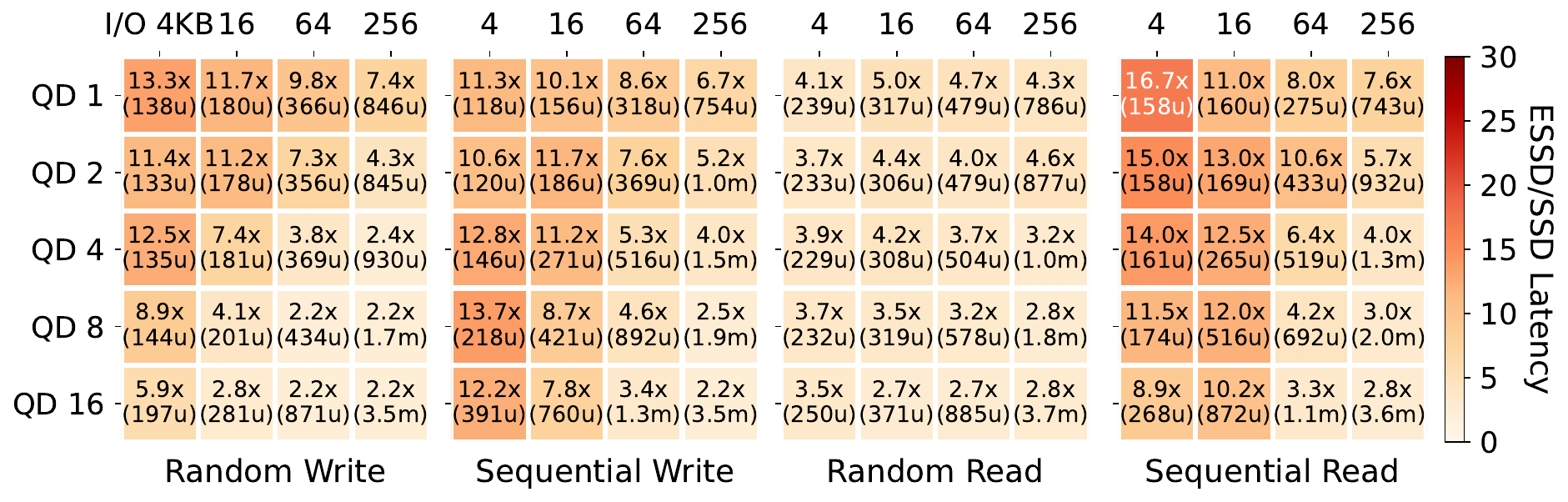}
    }
    \hspace{-2.1em}
    \subfigure[P99.9 Latency of ESSD-2 from Alibaba Cloud]{
    \includegraphics[width=0.5\textwidth]{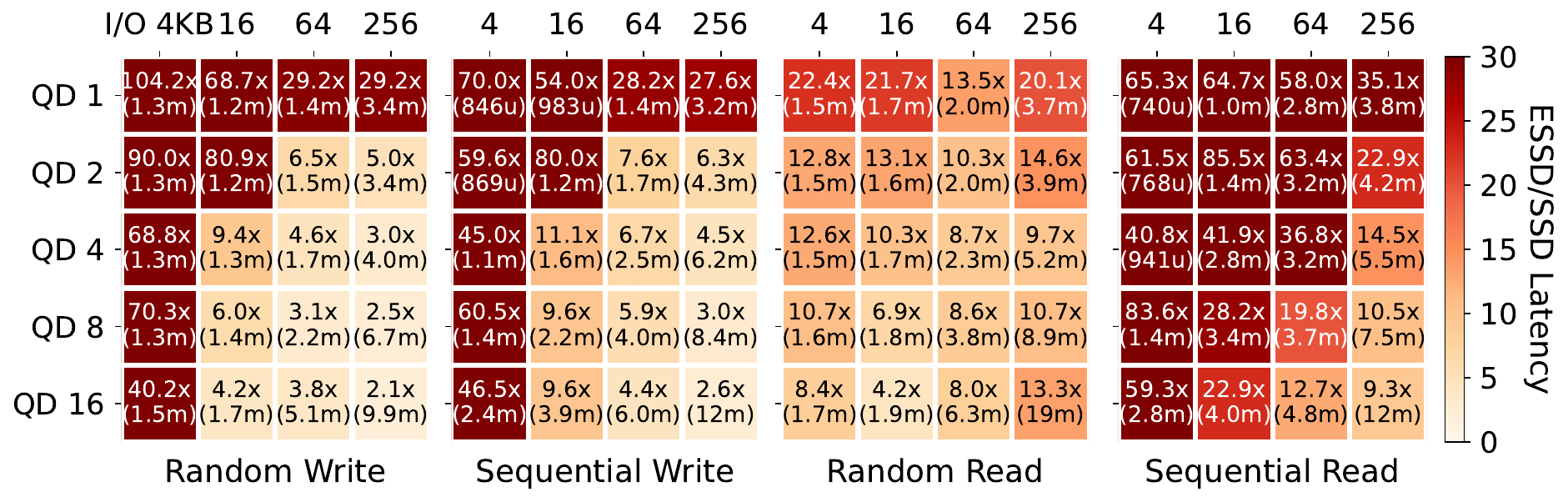}
    }
    \caption{
    \textbf{Latency performance of two ESSDs under workloads with different access patterns (i.e., random write, sequential write, random read, sequential read), I/O sizes, and I/O queue depths (QD).}
    The top of each pixel indicates the multiples the ESSD latency is divided by the SSD latency, with smaller numbers being better.
    The bottom of each pixel indicates the value of average or P99.9 latency of the ESSD (u: $\mu$s, m: ms).
    }
    % \vspace{.5em}
    \label{exp_lat}
\end{figure*}

% ------------------------------------------
\vspace{.5em}
\noindent\fbox{\parbox{0.475\textwidth}{
\emph{\textbf{Observation \#1:} The latency of ESSDs is tens to a hundred times higher than that of SSD when I/Os are not well scaled up (i.e., I/O size is small and/or I/O queue depth is low).}
}}
\vspace{.3em}
% ------------------------------------------

ESSD can meet the user's best expectations if it can provide comparable performance (e.g., throughput and latency) to the local SSD.
However, even though the maximum throughput that ESSD can provide reaches the level of GBs, we discover that the latency of ESSDs is extremely unsatisfactory when I/Os are not well scaled up.
Here, we define \textit{"not well scaled up"} as workloads dominated by small-size I/Os and/or low I/O queue depths.
Figure~\ref{exp_lat} demonstrates the average and P99.9 latency of two ESSDs under different access patterns (i.e., random write, sequential write, random read, and sequential read) with I/O sizes varied from 4KB to 256KB and I/O queue depths varied from 1 to 16.
Particularly, let us focus on the multiples the ESSD latency is divided by the SSD latency (i.e., the top of each pixel in Figure~\ref{exp_lat}), with smaller numbers being better.
For simplicity, we call this metric \textit{latency gap} in the following of this subsection.

We can first observe that, the latency gap effectively decreases when I/Os are scaled in either sizes or queue depths.
This finding is generally consistent among two metrics (i.e., average and P99.9 latency), four access patterns (i.e., random write, sequential write, random read, and sequential read), and two ESSD providers (i.e., Amazon AWS and Alibaba Cloud).
Specifically, the average latency gaps of two ESSDs are up to 47.9$\times$ and 16.7$\times$, respectively, while the P99.9 latency gaps are even much higher at up to 99.4$\times$ and 104.2$\times$.
In comparison, when the I/O size is scaled up to 256KB, the average latency gaps of two ESSDs are reduced to up to 15.6$\times$ and 7.6$\times$, respectively, while the P99.9 latency gaps are mitigated to 20.9$\times$ and 35.1$\times$ at most.
This trend also holds true when scaling up the I/O queue depths.

The primary cause of the high latency of ESSDs would be the network latency and software processing overhead within the cloud storage, which is still substantial compared to the local storage.
Scaling the I/O sizes and I/O queue depths can effectively reduce the latency gap by leveraging parallelism in I/O processing and storage access distribution (see~\S\ref{sec:bg_essd}).

We can also notice that, the latency gaps in two ESSDs are significantly smaller in the random read workload but much higher in the other three workloads (i.e., random write, sequential write, and sequential read).
Taking ESSD-1 as an example, the average and P99.9 latency gaps are up to 9.3$\times$ and 12.8$\times$ in the random read workload, but as high as 31.9$\times$ and 36.5$\times$, 29.9$\times$ and 35.2$\times$, 53.2$\times$ and 99.4$\times$ in the other three workloads.

This performance discrepancy is because modern SSDs typically employ a DRAM-based write buffer to improve (both random/sequential) write performance and also use prefetching to enhance sequential read performance~\cite{li2022fantastic}.
Thus, in random/sequential write and sequential read workloads, there are fewer flash accesses on the critical I/O path.
However, the prefetching techniques work poorly for random reads, resulting in a larger number of flash accesses.
This would make the inherent overhead in cloud storage (e.g., network latency and software processing) less prominent than in the other three workloads, thereby leading to a smaller latency gap.

The above results and analysis suggest that, to save the cloud software from tens to a hundred times latency penalty, cloud storage users should revisit their software designs to \textit{scale the sizes and queue depths of I/Os up as much as possible} (\textbf{Implication \#1}).
Especially, when I/Os are well scaled up in random and sequential write workloads, ESSD-1 even achieves up to 3.3$\times$ better P99.9 latency compared to the local SSD (see Figure~\ref{exp_lat}b).

% ------------------------------------------
\subsection{Garbage Collection (GC)} \label{sec:uc_gc}
% ------------------------------------------

% ------------------------------------------
\vspace{.5em}
\noindent\fbox{\parbox{0.475\textwidth}{
\emph{\textbf{Observation \#2:} The performance impact of GC appears much later or even disappears.}
}}
\vspace{.3em}
% ------------------------------------------

\begin{figure}[t]
    \centering
    \includegraphics[width=0.49\textwidth]{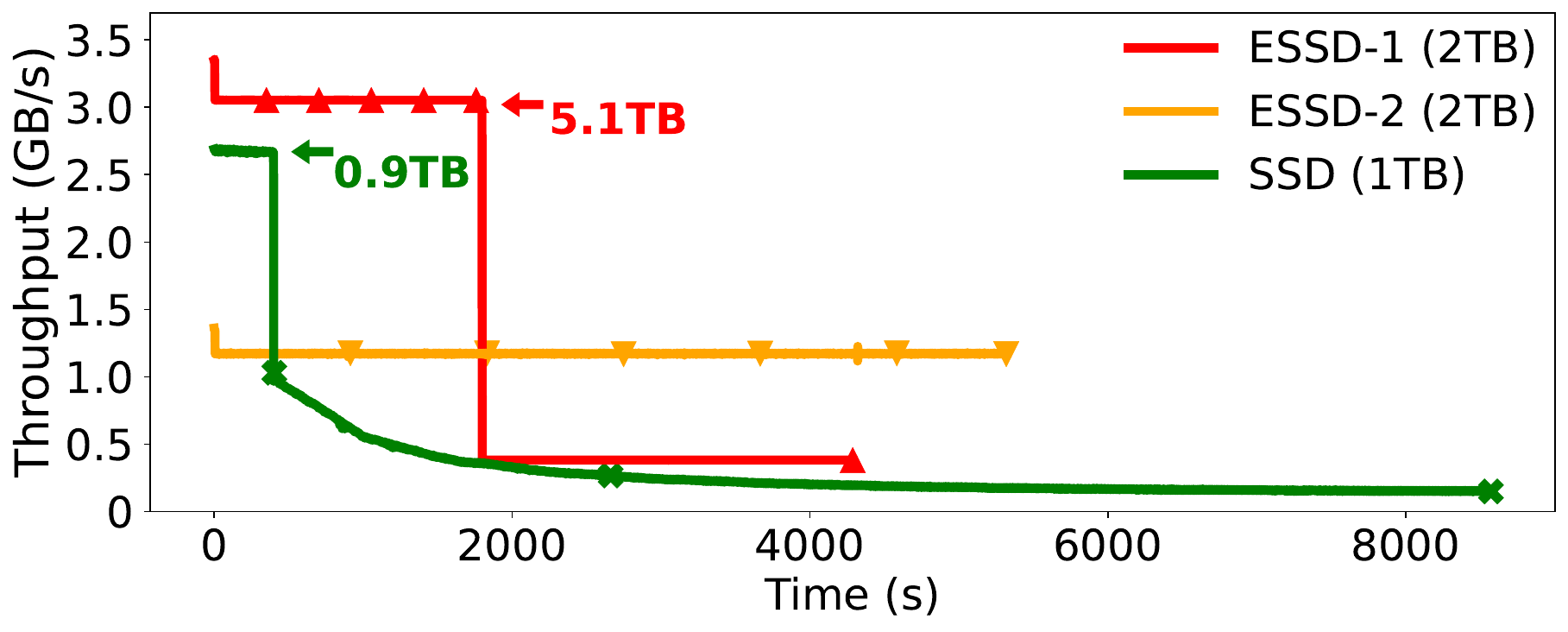}
    \vspace{-1.5em}
    \caption{
    \textbf{Runtime throughput of two ESSDs and SSD under a random write workload until writing 3$\times$ storage capacity.}
    Thus, the total write volumes towards ESSDs and SSD are 6TB and 3TB, respectively.
    The markers indicate the time when the total write volume reaches a multiple of 1TB.
    }
    % \vspace{-1em}
    \label{exp_gc}
\end{figure}

Device-side garbage collection (GC) is widely recognized as one of the main culprits for SSD performance degradation and unpredictability~\cite{yan2017tiny,elyasi2019trimming,li2021ioda}.
The main reason behind this is that, to reclaim the space occupied by the invalid data, FTL periodically relocates the valid data in some flash blocks to others.
The relocation of data results in a large number of extra writes, which severely compete with foreground I/Os and thus degrade the user-perceived performance.
Nevertheless, in contrast to this common anticipation, we find that the performance impact of GC appears much later or even disappears in ESSDs.
Figure~\ref{exp_gc} depicts the runtime throughput of two ESSDs and SSD under a random write workload, and the total write volume is configured to be 3$\times$ storage capacity of each device.

We can first verify that, the SSD starts to experience a severe throughput drop when only writing 0.9TB (i.e., 90\% storage capacity), and the throughput sharply decreases by 63\% from 2.7GB/s to 1.0GB/s.
After that, the throughput of SSD continues to decrease further as the write volume increases, down to a low of 149MB/s finally.
The long-term low performance here demonstrates the sensitivity and powerlessness of the SSD performance to GC.
Due to the high and constant write load, GC is unavoidably executed in the SSD almost all the time.

\begin{figure*}[t]
    \centering
    \includegraphics[width=1.01\textwidth]{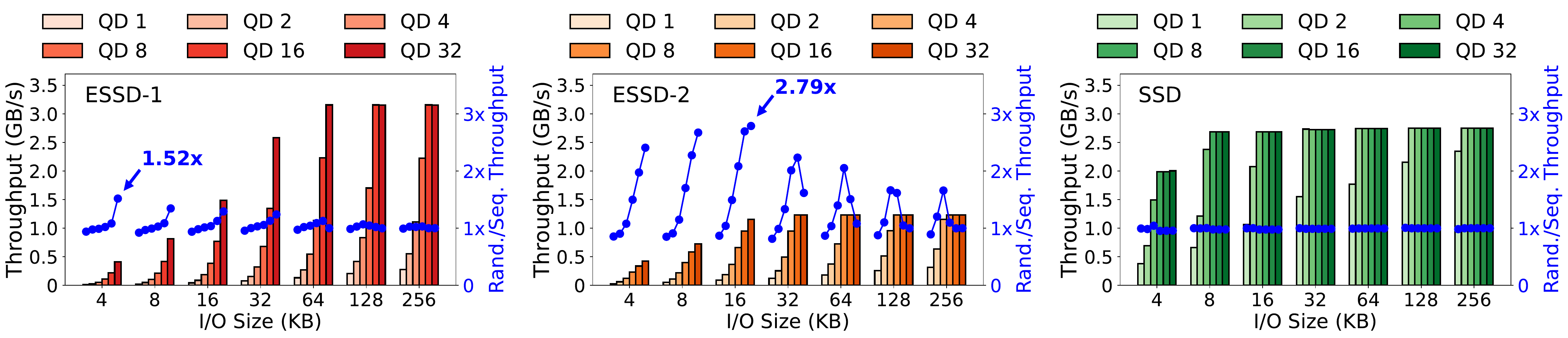}
    \vspace{-1.5em}
    \caption{
    \textbf{Throughput of ESSDs and SSD under random write workloads and throughput gain of ESSDs and SSD in random writes over sequential writes.}
    The I/O size varies from 4KB to 256KB and the I/O queue depth (QD) varies from 1 to 32.
    }
    % \vspace{.5em}
    \label{exp_bw}
\end{figure*}

In contrast, ESSD can sustain high throughput over a much longer period even in this high-pressure workload, although the exact performance of an ESSD may depend on the provider.
Specifically, ESSD-1 undergoes a sharp throughput decline when writing as large as 5.1TB (i.e., 2.55$\times$ storage capacity), after which the throughput is stabilized at about 305MB/s.
On the contrary, ESSD-2 can maintain a high throughput consistently until writing 6TB (i.e., 3$\times$ storage capacity).
We speculate on the rationale behind this as follows.
On the one hand, cloud providers make efforts to utilize their plenty of storage resources to hide the GC impact transparently for better user experiences.
On the other hand, cloud providers may trigger flow-limiting mechanisms when they can not hide the GC impact anymore.

Actually, it has been a long journey for researchers to study software techniques that can mitigate the performance impact of GC~\cite{skourtis2014flash,kim2019alleviating,jiang2021fusionraid}.
However, our results and analysis indicate that, cloud software should \textit{reconsider if and how existing GC-mitigated techniques based on local SSDs should be correspondingly adapted for ESSDs} (\textbf{Implication \#2}), due to the trade-offs they commonly entail.

% ------------------------------------------
\subsection{Access Pattern} \label{sec:uc_acp}
% ------------------------------------------

% \begin{figure}[t]
%     \centering
%     \subfigure[Throughput of ESSD-1 from Amazon AWS]{
%     \includegraphics[width=0.45\textwidth]{exp/exp_bw_essd1.pdf}
%     }
%     \subfigure[Throughput of ESSD-2 from Alibaba Cloud]{
%     \includegraphics[width=0.45\textwidth]{exp/exp_bw_essd2.pdf}
%     }
%     \subfigure[Throughput of Samsung 970 Pro SSD]{
%     \includegraphics[width=0.45\textwidth]{exp/exp_bw_ssd.pdf}
%     }
%     \vspace{-.5em}
%     \caption{
%     \textbf{Throughput in random writes and throughput improvement of random writes than sequential writes of ESSDs and SSD with different I/O sizes and queue depths (QD).}
%     }
%     \label{exp_bw}
% \end{figure}

% ------------------------------------------
\vspace{.5em}
\noindent\fbox{\parbox{0.475\textwidth}{
\emph{\textbf{Observation \#3:} The throughput of random writes actually outperforms that of sequential writes, reaching a maximum of 1.52$\times$ and 2.79$\times$ in two ESSDs, respectively.}
}}
\vspace{.3em}
% ------------------------------------------

Random writes have long been considered harmful to the SSD due to the exacerbated mix of valid/invalid data in the same flash blocks, which consequently results in higher valid data relocation overhead~\cite{chen2009understanding,min2012sfs,he2017unwritten}.
% Since the performance of ESSD is quite insensitive to GC (see~\S\ref{sec:uc_gc}), the performance of random writes is expected to be close to that of sequential writes.
However, surprisingly, we discover that the throughput of the ESSD is generally better in the case of random rather than sequential writes.
As shown in Figure~\ref{exp_bw}, let us focus on the throughput gain of ESSDs and SSD in random writes over sequential writes (i.e., blue lines).
For simplicity, we call this metric \textit{throughput gain} in the following of this subsection.
The I/O size varies from 4KB to 256KB and the I/O queue depth varies from 1 to 32.

We can first observe that, the throughput gains in two ESSDs are up to 1.52$\times$ and 2.79$\times$, respectively.
On the contrary, the throughput of the SSD does not show obvious differences in random or sequential writes (when GC does not occur).
The main reason behind this would be the unique physical storage space mapping of ESSDs.
As introduced in~\S\ref{sec:bg_essd}, the storage space of an ESSD is distributed and replicated across different nodes and SSDs in the storage cluster.
Therefore, data from random writes is likely to embrace higher available SSD bandwidth, potentially leading to higher user-perceived throughput.

Additionally, we discover a notable variation in throughput gains amongst ESSD providers.
Specifically, ESSD-1 has a smaller throughput gain, which is mainly concentrated on workloads with higher I/O queue depths and small-to-medium I/O sizes.
For example, when the I/O size ranges from 4KB to 32KB and the I/O queue depth is 32, the throughput gain of ESSD-1 is 24.1\%$\sim$51.8\%.
In contrast, the throughput gain in ESSD-2 is much more significant, at up to 1.66$\times$ to 2.79$\times$ across a wide range of I/O sizes (i.e., 4KB to 256KB).
The throughput gain in ESSD-2 also varies a lot in different workload scenarios. 
For small-size I/Os (e.g., 4KB to 16KB), the throughput gain generally increases as the queue depth increases.
For medium-to-large I/Os (e.g., 32KB and larger), the throughput gain has a trend of growing and then declining as the queue depth increases; moreover, the peak of the throughput gain occurs earlier as the I/O size increases.

Converting random writes into sequential writes (e.g., log-structured~\cite{lee2015f2fs} or copy-on-write~\cite{rodeh2013btrfs}) has been widely recognized to reduce device-side GC overhead and improve SSD performance.
However, given the ESSD performance is quite insensitive to GC (see~\S\ref{sec:uc_gc}) and the performance gain from random writes, it is now essential to \textit{rethink if it is worthwhile to convert random writes in random-write-based software into sequential writes anymore and if it is beneficial to proactively trigger random writes in sequential-write-based software} (\textbf{Implication \#3}).

% ------------------------------------------
\subsection{Throughput Budget} \label{sec:uc_mix}
% ------------------------------------------

% ------------------------------------------
\vspace{.5em}
\noindent\fbox{\parbox{0.475\textwidth}{
\emph{\textbf{Observation \#4:} The maximum bandwidth is now deterministic and no longer sensitive to the access pattern.}
}}
\vspace{.3em}
% ------------------------------------------

\begin{figure}[t]
    \centering
    \includegraphics[width=0.49\textwidth]{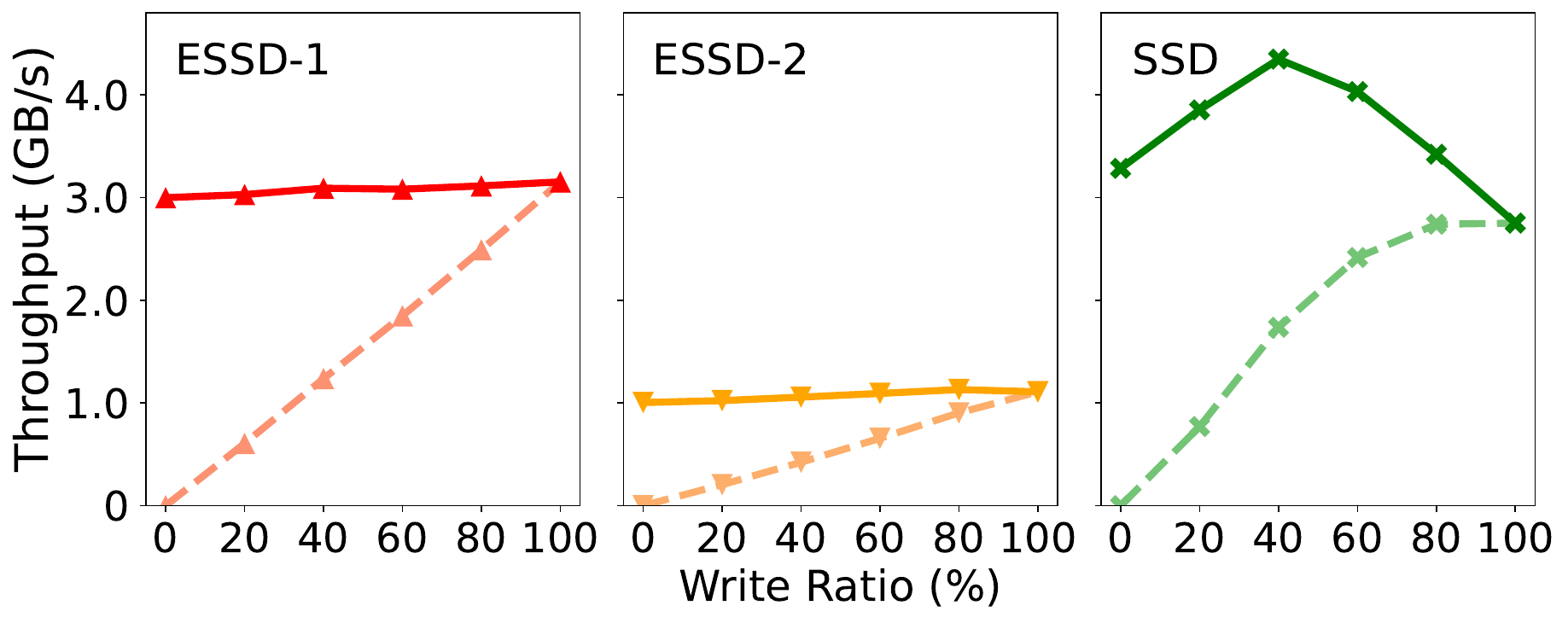}
    \vspace{-1.7em}
    \caption{
    \textbf{Throughput of two ESSDs and SSD under mixed read/write workloads with different write ratios.}
    The solid and dashed lines represent the total (i.e., read and write) throughput and the write throughput, respectively.
    }
    % \vspace{-1em}
    \label{exp_mix}
\end{figure}

The maximum throughput of the SSD largely depends on the access pattern, mainly due to the asymmetric latency of different flash operations.
For example, since flash reads typically have shorter latency than flash writes, the maximum read bandwidth is usually higher than the maximum write bandwidth (e.g., 3.5GB/s vs. 2.7GB/s in Samsung 970 Pro, see Table~\ref{essd_config}).
However, we find that the maximum bandwidth of ESSDs is deterministic and no longer sensitive to the access pattern.
Figure~\ref{exp_mix} shows the throughput of two ESSDs and SSD under mixed read/write workloads.
The workloads have different write ratios from 0 to 100, where 0 and 100 denote the random read and write workloads, respectively.

We can observe that, regardless of the specific write ratio, the throughput of two ESSDs is deterministically surrounded at the value guaranteed by the provider, i.e., 3.0GB/s and 1.1GB/s, respectively.
In contrast, the throughput of the SSD is non-deterministic, varying between 2.5GB/s and 4.3GB/s in different write ratios.
Note that this observation holds only for throughput and not for IOPS, as we find that the guaranteed IOPS in ESSDs is non-deterministic and is closely related to the I/O size.

The above results and analysis actually serve as a reminder to cloud storage users that, in contrast to the intricate performance engineering on the SSD, the throughput optimization target of the ESSD is straightforward and clear.
For cloud storage users, this can deliver two implications.

First, since the throughput budget directly affects the system cost (see~\S\ref{sec:bg_essd}), to achieve higher cost-effectiveness through a smaller throughput budget, cloud software should \textit{smooth the read/write I/Os to be evenly distributed across the timeline and below the guaranteed throughput budget} (\textbf{Implication \#4}).
This implication would be much more meaningful for software with frequent I/O bursts (i.e., I/O loads are very uneven across the timeline).

% % ------------------------------------------
% \vspace{.5em}
% \noindent\fbox{\parbox{0.475\textwidth}{
% \emph{\textbf{Observation \#5:} The throughput budget is budget-driven (elastic) -> for budget-limited, software help is important, e.g., reduce i/o amplification}
% }}
% \vspace{.3em}
% % ------------------------------------------

Second, to achieve higher cost-effectiveness, cloud software should also \textit{re-evaluate I/O reduction techniques (e.g., compression~\cite{zuck2014compression}, deduplication~\cite{xia2016comprehensive}) that were previously considered to impair performance} (\textbf{Implication \#5}).
The rationale behind this is that, for local SSDs with low latency, these I/O reduction techniques usually result in performance degradation due to high computational overhead~\cite{zuck2014compression,xia2016comprehensive}.
However, for ESSDs with network latency and software processing overhead that accompany cloud storage, the overhead of these techniques may not be the bottleneck, leading to not only cost reduction (i.e., accommodate workloads with a smaller throughput budget) but also performance improvement.

% ------------------------------------------
\vspace{1em}
\section{Related Work} \label{sec:related}
% ------------------------------------------

Depicting the performance characteristics of emerging storage devices is always a timeless topic.
Prior works have investigated the performance features of HDD\cite{schlosser2004mems}, flash-based SSD~\cite{he2017unwritten}, Optane-based SSD~\cite{wu2019towards}, and flash-based SSD with new interfaces such as Key-Value (KV)~\cite{saha2021kv} and Zoned Namespace (ZNS)~\cite{doekemeijer2023performance}.
Inspired by them, we first characterize the performance of the emerging cloud-based ESSD.

Some works introduce EBS's designs and evolution in storage, networking, and hardware acceleration~\cite{miao2022luna,zhang2024s,wang2024ransom,shu2024burstable}.
Li et al. characterize the I/O features of workloads on EBS from Alibaba Cloud and Tencent Cloud~\cite{li2023depth}.
Cosine~\cite{chatterjee2021cosine} searches for the best key-value storage engine with its cost models considering an input workload, a cloud budget, a target performance, and required cloud SLAs.
Many works focus on data management between EBS and other faster/slower tiers (e.g., local SSD, object store such as Amazon S3~\cite{website:s3}).
For example, Mutant~\cite{yoon2018mutant} proposes to organize different SSTables in the key-value store into different storage tiers based on access frequencies.
ROCKSMASH~\cite{xu2022building} proposes new cache and metadata structure designs to use local SSDs to store frequently accessed data and metadata while using EBS to hold the rest.
MirrorKV~\cite{wang2023mirrorkv} focuses on hybrid cloud storage consisting of EBS and object store, and proposes a series of designs to address the efficiency challenges of compaction and querying in this architecture.

Zhou et al. reveal the high latency variances in ESSDs when exceeding the paid IOPS budget and propose a range of I/O-regulated techniques in the key-value store to improve tail latency~\cite{zhou2023calcspar}.
In contrast, we focus on the comprehensive performance characterization of ESSDs and deliver insights from many aspects (e.g., latency performance, GC, access pattern, throughput budget).
% (i.e., not limited to the budget-exceeded latency performance).

% ------------------------------------------
% \vspace{1em}
\section{Conclusion and Future Work} \label{sec:conclusion}
% ------------------------------------------

This paper pioneers a thorough performance characterization of the cloud-based ESSD, and draws an unwritten contract that incorporates four counter-intuitive observations and five intriguing implications.
We hope the unwritten contract could guide cloud storage users in revisiting their deployed cloud software designs, particularly, harnessing the distinct performance characteristics of ESSDs for better system performance.

Meanwhile, given the increasingly rapid growth of cloud computing and cloud storage, we believe this research direction deserves more investigation.
We outline the future work as follows.

\begin{itemize}[leftmargin=*]
    \item \textbf{Comprehensiveness and generalizability.} 
    % To strengthen the generalizability of findings, we have evaluated two ESSDs from two leading cloud storage providers: Amazon AWS and Alibaba Cloud.
    We acknowledge that our current observations may not cover all the unique properties of ESSDs.
    In the future, we will continue to investigate, and also evaluate ESSDs from more cloud storage providers and validate our observations in this paper on these ESSDs.
    \item \textbf{Cloud-software-level exploration.}
    In the future, we will explore how cloud software can take advantage of the implications offered in this paper.
    In particular, we will first use RocksDB~\cite{website:rocksdb} (a popular persistent key-value store) as a case study.
    Although some works have explored building key-value stores on cloud storage (see~\S\ref{sec:related}), these works may not fully exploit the potential of ESSDs.
\end{itemize}

\newpage

\bibliographystyle{IEEEtran}
\bibliography{uc}

\end{document}